\documentclass[12pt]{article}

\usepackage{a4wide}
\usepackage[ngerman,english]{babel}
\useshorthands{"}
\addto\extrasenglish{\languageshorthands{ngerman}}
\usepackage[utf8]{inputenc}
\usepackage[T1]{fontenc}
\usepackage{amsmath}
\usepackage{graphicx}
\usepackage{amssymb}
\usepackage{hyperref}
\usepackage{mathrsfs}
\usepackage{color}

\renewcommand{\phi}{\varphi}
\renewcommand{\theta}{\vartheta}

\title{{Schwarzschild/CFT from soft black hole hair?}}
\author{{\bf Artem Averin$^{\textrm{a,b}}$}}

\begin{document}

\maketitle

\centerline{\it $^{\textrm{a}}$ Arnold--Sommerfeld--Center for Theoretical Physics,}
\centerline{\it Ludwig--Maximilians--Universit\"at, 80333 M\"unchen, Germany}
\medskip
\centerline{\it $^{\textrm{b}}$ Max--Planck--Institut f\"ur Physik,
Werner--Heisenberg--Institut,}
\centerline{\it 80805 M\"unchen, Germany}

\vskip1cm
\abstract{{
Recent studies of asymptotic symmetries suggest, that a Hamiltonian phase space analysis in gravitational theories might be able to account for black hole microstates. In this context we explain, why the use of conventional Bondi fall-off conditions for the gravitational field is too restrictive in the presence of an event horizon. This implies an \emph{enhancement} of physical degrees of freedom ($\mathcal{A}$-modes). They provide new gravitational hair and are responsible for black hole microstates. Using covariant phase space methods, for the example of a Schwarzschild black hole, we give a proposal for the surface degrees of freedom and their surface charge algebra. The obtained two-dimensional dual theory is conjectured to be conformally invariant as motivated from the criticality of the black hole. Carlip's approach to entropy counting reemerges as a Sugawara-construction of a 2D stress-tensor.  
}}

\begin{flushright}
\end{flushright}

\newpage

\setcounter{tocdepth}{2}
\tableofcontents
\break

\section{Introduction}

\subsection{The information paradox for black holes}

One of the most robust predictions of quantum gravity is that black hole formation is accompanied by its subsequent evaporation via Hawking radiation \cite{Hawking}. Hawking's calculation predicts that this radiation has a unique thermal spectrum. This observation leads to the information paradox: Letting the black hole evaporate and observing its radiation, it seems as a matter of principle impossible to retrieve information about how the black hole was formed. Unitarity seems to be violated (see \cite{Carlip:2014pma} for a review).

Hawking's calculation is done by treating the background metric as a classical field (on top of which additional fields are quantized). This approximation receives of course corrections and it was proposed in \cite{Dvali:2011aa} \cite{Dvali:2012rt} \cite{Dvali:2013eja} \cite{Dvali:2012wq} that they are sufficient to resolve the paradox.

In an arbitrary quantum field theory, there can be quantum states, in which the approximation of working with classical fields and using classical equations of motion is a good approximation (also known as the mean-field approximation in several contexts). This approximation receives corrections which are suppressed by a factor of some power of $(\text{\it{size of system}})^{-1}.$ Remembering the analogy of quantum field theory and statistical mechanics, they are the analog of the statistical fluctuations of an observable around its expectation value in an ensemble. These fluctuations are also suppressed by some power of $(\text{\it{size of system}})^{-1}.$ In \cite{Dvali:2011aa} \cite{Dvali:2012rt} \cite{Dvali:2013eja} \cite{Dvali:2012wq} these corrections were termed $\frac{1}{N}$-corrections ($N$ being a parameter describing the size of the system) and their meaning for the Hawking-effect was stressed. 

The thermal spectrum of emitted quanta gets corrected by these $\frac{1}{S}$-effects (the size $N$ can be measured by the black hole entropy $S$). These corrections provide observable features from which (in principle) the information can be retrieved how the black hole was formed. After the half-life time of the black hole the $\frac{1}{S}$-corrections accumulate, so that the spectrum is far from thermality and information recovery starts to get efficient in accordance with Page's time \cite{Page:1993wv}. Ignoring $\frac{1}{S}$-corrections (this is the limit in which Hawking's calculation is performed), one is left with the information paradox. 

However, even if the Hawking spectrum is corrected by $\frac{1}{S}$-effects, the different $\frac{1}{S}$-effects must be sourced by different black hole microstates in order to be able to contain information about black hole formation. In other words, there must be a huge number of states in the Hilbert space, that correspond to the microstates of a suited black hole in agreement with the Bekenstein-Hawking entropy \cite{Bekenstein:1973ur}. In pure Einstein gravity, the entropy is infinite in the classical ($\hbar \to 0$) limit. Thus, in the Hamiltonian phase space, there must be an infinite number of points corresponding to the microstates of a particular black hole. Where are these points in phase space? This is the question, that will be the subject of our investigations.

\subsection{Kerr/CFT from Criticality}

That black hole microstates have to be visible in the Hamiltonian phase space of Einstein gravity can be motivated also from another direction. In \cite{Dvali:2012en} the appearance of microstates and thus of black hole entropy is explained as to have its physical origin due to the following general field-theoretic phenomenon:

Suppose a theory with a bosonic field and attractive self-interaction. A field-configuration, which is right at the point of being self-sustained, that is, to be stationary and localized in space by its own attractive self-interaction, is accompanied by the appearance of gapless excitation modes.

The latter point is called a critical point and gapless here is meant with respect to the classical Hamiltonian (i.e. degeneracy in energy). Examples of this phenomenon are well-known in much simpler field theories from condensed matter physics (see \cite{Dvali:2012en} \cite{Flassig:2012re} \cite{Dvali:2013vxa} and references therein). The excitation modes of such field-configurations are in several contexts also called Bogoliubov-modes. The critical point described is thus accompanied with the appearance of gapless Bogoliubov-modes. The degeneracy is in the quantum theory lifted by $\frac{1}{N}$-effects. This implies that $\frac{1}{N}$-corrections can accumulate over a time-scale set by the size of the system $N$ and deviate significantly with the predictions of a mean-field analysis. Therefore, quantum corrections can not be neglected at a critical point (even if the system is large). The critical point is a quantum critical point. 

Now, the case of pure Einstein gravity provides a special case to what we have said. The stationary, asymptotically flat solutions are given by the Kerr-family \cite{Heusler}. These are critical field configurations and as such must possess gapless Bogoliubov-modes. These gapless modes are the physical origin of the black hole entropy. Thus, the Hamiltonian phase space $\Gamma$ of Einstein gravity has to contain a region $S \subseteq \Gamma$ containing the Kerr-family and its gapless Bogoliubov-excitations. 

Due to this scale-invariance, it is tempting to expect that the part $S$ of Hamiltonian phase space has a conformal invariance. The lifting of the degeneracy of the Bogoliubov-modes by the $\frac{1}{S}$-effects in the quantum theory is reflected by conformal anomaly of this invariance. 

Indeed, such a Kerr/CFT-correspondence was proposed \cite{Castro:2010fd} as an extrapolation of the extremal Kerr/CFT-correspondence \cite{Guica:2008mu}. By analysis of scattering off a non-extremal Kerr black hole, some data of the dual CFT could be obtained and were shown to be in agreement with Bekenstein-Hawking entropy. However, neither a formulation of the dual CFT has so far been obtained nor was it understood why there is a Kerr/CFT-correspondence. The physical origin of a possible Kerr/CFT-correspondence, we explain with the criticality of the Kerr solutions.

\subsection{Kerr/CFT from soft black hole hair}

We have argued from various directions that the Hamiltonian phase space of pure Einstein gravity has to contain an infinite number of gapless excitations of the Kerr-family. But then, there is a problem. Where are these excitations that are among other things responsible for black hole microstates? According to the black hole uniqueness theorems, all solutions of the field equations that are asymptotically flat and stationary are given by the Kerr-metric up to diffeomorphisms. The hope is then that not all of these diffeomorphisms are gauge redundancies. Some of them should be physical excitations, i.e. shifts in the phase space, providing the necessary gapless excitation modes. This idea goes for the case of four-dimensional black holes already back to Carlip \cite{Carlip:1998wz}, has later on be one of the main motivations in the study of asymptotic symmetries \cite{Barnich:2010eb} and has recently gained attention as the soft hair on black holes proposal \cite{Hawking:2016msc}. However, a satisfactory analysis of the phase space so far has not been given in the literature.

In this paper, we want to make a first step in this direction. Using mainly covariant phase space methods, we want to analyze the phase space near a Schwarzschild black hole solution. More specifically, we look at its gauge excitations and single out its surface degrees of freedom (chapter 3). These are found to violate the conventionally used Bondi fall-off conditions for the gravitational field. We explain in general, why these fall-off conditions are too restrictive in the presence of an event horizon (chapter 2). Calculating the surface charge algebra (chapter 4), we are able to propose a two-dimensional theory for the surface degrees of freedom of a Schwarzschild black hole. Remarkable is the appearance of central terms which supports the conjecture that the dual theory is a CFT (Schwarzschild/CFT-correspondence).

We want to warn that the present work is just a first step and there are still a lot of things to be understood. An analysis of the phase space structure in the region of the Kerr-family is beyond our present scope. However, we explain which assumptions entered in the derivation of our dual theory (chapter 5) and give an outlook what is at our current investigation. Especially, Carlip's approach to entropy counting is in our approach a Sugawara-construction of a 2D stress-energy tensor for our dual theory. It is then tempting to expect that this dual theory is a CFT describing the phase space of the whole Kerr-solutions (Kerr/CFT-correspondence).

In the following, we use units in which we set the speed of light to $1$ but we keep Newton's constant $G$ and Planck's constant $\hbar$ explicit. Latin letters $a, b, \ldots = 0, \ldots, 3$ denote spacetime indices.  

\section{Cauchy-Data for asymptotically flat 4d spacetimes}

We start by asking what is a possible set of Cauchy-data (gauge-fixed solution space) to specify a solution describing a particular state in phase space in Einstein gravity? This question already appeared in the study of gravitational waves and we adopt the answer which is reviewed in \cite{Barnich:2010eb}. We denote coordinates by $(x^0, x^1, x^A)=(u,r, \theta , \phi)$ with $A,B, \ldots = 2,3.$ The Bondi gauge-fixing conditions read

\begin{equation} \label{1}
\begin{split}
g_{rr}=g_{rA}=0 \\
\det g_{AB} = r^4 \sin^2 \theta.
\end{split}
\end{equation}

Imposing Bondi fall-off conditions, the metric is written as

\begin{equation}
ds^2 = e^{2\beta} \frac{V}{r} du^2 - 2e^{2\beta} du dr + g_{AB}\left (dx^A-U^Adu\right ) \left (dx^B-U^Bdu \right )
\label{2}
\end{equation}

with

\begin{equation}
g_{AB}dx^A dx^B = r^2 \gamma_{AB} dx^A dx^{B} + O(r),
\label{3}
\end{equation}

where

\begin{equation}
\gamma_{AB} dx^A dx^B = d\theta^2 + \sin^2 \theta d\phi^2
\label{4}
\end{equation}

is the metric on the unit $2$-sphere. The remaining fall-off conditions are 

\begin{equation} \label{5}
\begin{split}
\beta &= O(r^{-2}) \\
\frac{V}{r} &= -1 + O(r^{-1}) \\
U^A &= O(r^{-2}).
\end{split}
\end{equation}

The Bondi-gauge with required fall-offs is suited to describe the gravitational field of asymptotically flat spacetimes near null infinity $J.$ (In this chapter, the required fall-offs correspond to retarded Bondi-gauge and cover the region near future null infinity $J^+.$)

A metric in Bondi-gauge and with Bondi fall-off conditions that is further satisfying vacuum Einstein field equations is fully determined by the set of functions 

\begin{equation}
\begin{split}
\mathcal{X} = \{
&N_{AB}(u,x^C); M(u_0,x^A); N_{A}(u_0,x^B); C_{AB}(u_0,x^C); \\
&D_{AB}(x^C); E_{AB}(u_0,r,x^C) \},
\end{split}
\label{6}
\end{equation}

for a fixed retarded time $u_0.$ That means, to specify a concrete solution, one has to specify the Bondi-News $N_{AB}(u,x^C),$ which characterize the gravitational radiation passing through null infinity. The remaining part of the Cauchy-data consists of functions on $S^2,$ which we will collectively denote boundary Cauchy-data (BCD). Among these are the mass and angular momentum aspects $M(u_0,x^A), \, N_A (u_0, x^B)$ for fixed time, as well as leading BCD $C_{AB}(u_0, x^C), \, D_{AB}(x^C)$ and subleading (in $r$) BCD summarized in the function $E_{AB}(u_0,r,x^C).$ 

For the conditions on the functions appearing in $\mathcal{X}$ and their connections to the metric \eqref{2}, we refer to \cite{Barnich:2010eb}. 

We point out that the Bondi fall-off conditions are also imposed in the determination of the asymptotic symmetry algebra. That means, the asymptotic symmetries are defined as the residual gauge transformations preserving the Bondi gauge-fixing \eqref{1} as well as Bondi fall-offs \eqref{2}-\eqref{5}. This results in the $\mathfrak{bms}_4$-algebra (see \cite{bms} \cite{Barnich:2009se} \cite{Barnich:2010eb} \cite{Barnich:2011mi} for the various definitions and realization on gauge-fixed solution space \eqref{6}). However, our point is that in the presence of an event horizon the Bondi fall-offs \eqref{2}-\eqref{5} are too restrictive. As a consequence, precisely in the presence of a black hole, there is an \emph{enhancement} in \eqref{6} in the required Cauchy-data by additional BCD.

As is already evident from the derivation of the gauge-fixed solution space \eqref{6} in \cite{Barnich:2010eb}, after relaxing the Bondi fall-offs, there are solutions with additional terms in \eqref{2} violating Bondi fall-offs. However, gravitational radiation passing through $J^+$ as characterized by the Bondi-News $N_{AB}$ has no effect on them. In other words, there is no associated memory effect. Any additional Cauchy-data is seen as a redundancy. 

The situation is different in the presence of an event horizon. There is \emph{a priory} the possibility, that gravitational radiation passing the event horizon can leave an imprint on the additional terms in \eqref{2} that violate the Bondi fall-offs. This is the possibility that we want to advocate here. The additional BCD labels the different black hole microstates. Choosing different BCD corresponds to exciting different microstates. Imposing Bondi fall-offs (and thus ignoring the additional BCD), one encounters a sort of black hole information paradox: Looking at the solution space \eqref{6}, there is no space for the black hole microstates. 

What is then the additional BCD that has to be included in \eqref{6} in the presence of an event horizon? In the next chapter, we try to answer this question for the example of a Schwarzschild black hole, in which case the data \eqref{6} reads

\begin{equation} \label{7}
\begin{split}
N_{AB} = C_{AB} &= D_{AB} = E_{AB} = 0 \\
N_{A} &= 0 \\
M &= \frac{r_S}{2G},
\end{split}
\end{equation}

where $r_S$ is the Schwarzschild radius. 

\section{Surface degrees of freedom of a Schwarzschild black hole}

The well-known Schwarzschild-metric is in Schwarzschild-coordinates given by

\begin{equation}
ds^2=-\left(1-\frac{r_S}{r}\right)dt^2+\left(1-\frac{r_S}{r}\right)^{-1}dr^2+r^2(d\theta^2+\sin^2 \theta d\phi^2).
\label{8}
\end{equation}

Defining the tortoise coordinate

\begin{equation}
r^*=r+r_S \ln \left| \frac{r}{r_S} - 1 \right|,
\label{9}
\end{equation}

one has

\begin{equation}
\frac{dr^*}{dr} = \left(1-\frac{r_S}{r}\right)^{-1}.
\label{10}
\end{equation}

Choosing ingoing Eddington-Finkelstein coordinates $(v,r,\theta,\phi)$ with

\begin{equation}
v=t+r^*,
\label{11}
\end{equation}

the metric reads 

\begin{equation} \label{12}
\begin{split}
ds^2 &= -\left(1-\frac{r_S}{r}\right)dv^2+2dvdr+r^2(d\theta^2+\sin^2 \theta d \phi^2) \\
&= -\left(1-\frac{r_S}{r}\right)dv^2+2dvdr+r^2 \gamma_{AB}dx^Adx^B.
\end{split}
\end{equation}

In this coordinates, the metric satisfies the Bondi-gauge conditions. However, note that from now on, we are working in advanced Bondi-gauge, in which the $r \to \infty$ limit, describes the region near past null infinity $J^-.$ 

We now fix a Schwarzschild-radius $r_S,$ then \eqref{12} provides a concrete reference point $g_{ab}$ in gauge-fixed solution space. Our task in this chapter is to find nearby points in (gauge-fixed) solution space $g_{ab}+h_{ab},$ which are candidates for the microstates of this particular Schwarzschild black hole with mass parameter $\frac{r_S}{2G}.$ As already noted at the end oft the last chapter, $g_{ab}+h_{ab}$ has to satisfy Bondi gauge-fixing conditions, but we expect it to violate Bondi fall-offs. 

How do we then find the relevant excitations $h_{ab}$ responsible for black hole microstates? 

Our strategy is that we insist on the existence of a consistent Hamiltonian description of the phase space at least in the neighboorhood of $g_{ab}.$ To analyze the Hamiltonian phase space near $g_{ab},$ we use covariant phase space methods \cite{Lee:1990nz} \cite{Wald:1999wa} although at some points the direct Hamiltonian approach is employed. A review of the covariant phase space approach (including the relevant references) is given in \cite{Seraj:2016cym}, whereas the Hamiltonian approach is reviewed in \cite{Troessaert:2013fba}. 

A helpful observation comes from the black hole uniqueness theorems, which state that every asymptotically flat and stationary solution of the vacuum field equations in four dimensions is \emph{diffeomorphic} to the Kerr-solution. Therefore, there is the possibility that the black hole microstates could be hidden in the form of excitations $h_{ab} = \mathcal{L}_{\xi} g_{ab}$ which take the form of gauge transformations. Of course, most of these excitations will correspond to gauge redundancies. However, there could be a subclass corresponding to the excitations of real physical degrees of freedom, i.e. a shift in the Hamiltonian phase space. This possibility was recently proposed in \cite{Hawking:2016msc} and termed ``soft hair on black holes.'' (See also the earlier work of Carlip \cite{Carlip:1998wz}.) Nevertheless, a determination of the relevant degrees of freedom responsible for microstates is still missing. We want to make a proposal in this direction. 

As explained, the candidate excitations $h_{ab}$ should preserve Bondi-gauge \eqref{1} and must take the form of a gauge transformation $h_{ab} = \mathcal{L}_{\zeta} g_{ab}$ for a vectorfield $\zeta.$ However, we do not impose any fall-off conditions. These residual gauge transformations are found to be \cite{Barnich:2010eb} \cite{Hawking:2016sgy} 

\begin{equation}
\zeta=\zeta \left(X,X^A \right) = X\partial_v -\frac{1}{2}\left(r D_A X^A+D^2X\right) \partial_r +\left(X^A+\frac{1}{r}D^AX\right)\partial_A.
\label{13}
\end{equation}

Here, $X=X(v,x^A)$ is an arbitrary scalar and $X^A=X^A(v,x^B)$ an arbitrary vectorfield on $S^2.$ Indices $A,B,\ldots = 2,3$ labeling coordinates on the sphere are raised and lowered with $\gamma_{AB}.$ $D_A$ denotes the associated covariant derivative and $D^2$ the Laplace-operator. The corresponding non-zero shifts in the metric components $h_{ab}=\mathcal{L}_{\zeta} g_{ab}$ are (with $V := 1-\frac{r_S}{r}$)

\begin{equation} \label{14}
\begin{split}
h_{vv} &= \frac{GM}{r}D_B X^B + \frac{GM}{r^2}D^2X-2V\partial_v X -r\partial_v D_BX^B-D^2\partial_vX \\
h_{Av} &= -\frac{r}{2}D_A D_B X^B -\frac{1}{2}D_AD^2X -VD_A X +r^2\partial_v X_A +r\partial_v D_A X \\
h_{AB} &= r^2(D_A X_B + D_B X_A -\gamma_{AB} D_C X^C) +r(2D_A D_B X - \gamma_{AB} D^2 X) \\
h_{vr} &= -\frac{1}{2} D_B X^B + \partial_v X.
\end{split}
\end{equation}

To investigate, which of the excitations \eqref{14} are physical, we inspect the Hamiltonian generators of these excitations. The relevant formulas of the covariant phase space approach are reviewed in \cite{Seraj:2016cym} on which we refer. We use also some formulas summarized in \cite{Hawking:2016sgy}. The covariant phase space $\bar{\mathcal{F}}$ is given by the (not gauge-fixed) solution space of the theory (set of field configurations satisfying equations of motion). After gauge-fixing, we obtain the gauge-fixed solution space $\Gamma,$ which can be taken up to residual symplectic zero-modes as the phase space. Since we are only interested in the gauge excitations of a Schwarzschild black hole, we will consider the fixed point $g_{ab} \in \Gamma$ and gauge excitations in the tangent space $T_{g_{ab}} \Gamma.$ In general, the Hamiltonian generator $H$ of a gauge transformation $\mathcal{L}_{\xi} g_{ab}$ over a Cauchy-surface $\Sigma$ is determined by 

\begin{equation}
\delta H[h_{ab};g_{ab}] = \int_{\Sigma} {\omega[h_{ab},\mathcal{L}_{\xi} g_{ab};g_{ab}]},
\label{15}
\end{equation}

where $\delta H$ denotes the variation of $H$ between the points $g_{ab}$ and $g_{ab}+h_{ab}.$ On-shell \eqref{15} reduces to a boundary integral    
  
\begin{equation}
\delta H[h_{ab};g_{ab}] = -\frac{1}{16 \pi G} \oint_{\partial \Sigma} {*F},
\label{16}
\end{equation}

for a well-known 2-Form $F$ over the spacetime. We will consider the expression \eqref{16}, where $\partial \Sigma$ is a cross-section from the event horizon. Thus $\partial \Sigma$ has fixed $v$ and $r=r_S$ and has the topology of an $S^2$ parametrized by the remaining coordinates $x^A.$ In this case, we have

\begin{equation}
\delta H[h_{ab};g_{ab}] = -\frac{r_S^2}{16 \pi G} \oint_{\partial \Sigma} {d^2x \sqrt{\gamma} F_{rv}},
\label{17}
\end{equation}

where $\gamma = \det \gamma_{AB}$ and

\begin{equation} \label{18}
\begin{split}
\left. F_{rv} \right|_{r=r_S} &= \xi^A\left(\partial_r h_{Av} -\frac{2}{r_S}h_{Av}\right) +\xi^v \left(-\frac{1}{r_S^2}D^A h_{Av} + \frac{1}{r_S^2}\partial_v {h^A}_A - \frac{2}{r_S}h_{vv} \right. \\
&- \left. \frac{1}{2r_S^3}{h^A}_A\right)
+ \partial_r \xi^v h_{vv} +\frac{1}{r_S^2}D^A\xi^v h_{vA} -\frac{1}{2r_S^2}\partial_v \xi^v {h^A}_A + \xi^r \left(\frac{2}{r_S}h_{vr} \right. \\
&+ \left. \frac{1}{r_S^3}{h^A}_A\right)
+ \frac{1}{2r_S^2}\partial_r \xi^r {h^A}_A.
\end{split}
\end{equation}

Here, the vectorfield $\xi$ is the gauge transformation to be implemented by $H$ and $h_{ab}$ satisfies linearized field equations around the fixed $g_{ab}$ but for later purposes $h_{ab}$ need not to be gauge fixed in \eqref{18}. (Therefore, \eqref{18} contains terms which vanish for $h_{ab}$ respecting Bondi-gauge.)

The change of the Hamiltonian generator $\delta H_{(Y,Y^A)}$ implementing a gauge excitation $\zeta = \zeta(Y,Y^A)$ (see \eqref{13}) under a gauge excitation $h_{ab}=h_{ab}(X,X^A)$ (see \eqref{14}) is then given by

\begin{equation} \label{19}
\begin{split}
&\delta H_{(Y,Y^A)} [h_{ab};g_{ab}] \\
&= \frac{r_S}{16 \pi G} \oint_{\partial \Sigma} {d^2x \sqrt{\gamma} \left( Y(1-D^2)D_B X^B + D_B Y^B (D^2-1)X \right)}.
\end{split}
\end{equation}

From \eqref{19}, we infer that excitations with

\begin{equation} \label{20}
\begin{split}
X &= X(x^A) \\
X^A &= X^A(x^B)
\end{split}
\end{equation}

with non-vanishing divergence $D_A X^A$ change the on-shell values of the Hamiltonian generators \eqref{19}.\footnote{Note that the differential operator $D^2-1$ is invertible on $S^2$ as it has no zero eigenvalues.} They are non-zero modes of the presymplectic form and thus constitute physical excitations of the Schwarzschild black hole. Furthermore, we see that any $v$-dependence which would be allowed in the residual gauge transformation \eqref{13} does not constitute any new physical excitation other than \eqref{20}.\footnote{Note that all dependence on $v$-derivatives of $X$ and $X^A$ cancels in \eqref{19}. \eqref{19} depends only on $X=X(v_0,x^A)$ and $X^A=X^A(v_0,x^B)$ with $v_0$ being the retarded time of $\partial \Sigma.$} At least from the point of view of the generators \eqref{19}, all physical gauge excitations of $g_{ab}$ are given by \eqref{20}. In other words, the physical gauge excitations (which form a subspace of $T_{g_{ab}} \Gamma$) can be parametrized (in Bondi-gauge) by the coordinates

\begin{equation} \label{21}
\begin{split}
X = &X(x^A) \\
D_A &X^A,
\end{split}
\end{equation}

where $X=X(x^A)$ is a scalar on $S^2$ and $X^A = X^A(x^B)$ is a vectorfield on $S^2.$ These excitations are physical in the sense that they are shifts in the phase space. They form the gauge or surface degrees of freedom of the Schwarzschild black hole. We will refer to the coordinates \eqref{21} as the \emph{gauge aspects.}

After having identified the gauge degrees of freedom of a Schwarzschild black hole \eqref{21}, we make some comments on their geometry and physics.

The choice

\begin{equation} \label{22}
\begin{split}
X &= f(x^A) \\
X^A &= 0
\end{split}
\end{equation}

for a function $f$ on $S^2$ in \eqref{13} corresponds to the usual $\mathfrak{bms}_4$-supertranslations \cite{Barnich:2010eb}. These excitations respect Bondi fall-offs and are thus contained in the (gauge-fixed) solution-space spanned by the Cauchy-data \eqref{6}. As explained in the last chapter, $\mathfrak{bms}_4$-supertranlations are thus not expected to be responsible for black hole microstates. Indeed, they just reflect the degeneracy of the gravitational vacuum \cite{Strominger:2013jfa}. It was already stated in \cite{Averin:2016ybl} that ordinary $\mathfrak{bms}_4$-supertranslations are not responsible for the microstates of a Schwarzschild black hole, but instead there is an enhanced asymptotic symmetry algebra. It is the \emph{enhancement} (which were called $\mathcal{A}$-modes), which were proposed to be responsible for the microstates and correct entropy counting \cite{Averin:2016hhm}. This reasoning resolves the criticism on the soft hair proposal correctly stated in \cite{Mirbabayi:2016axw}. 

What are the additional excitations contained in \eqref{21} besides \\
$\mathfrak{bms}_4$-supertranslations? For the vectorfield $X^A$ on $S^2$ we have a Helmholtz theorem, i.e. we can decompose

\begin{equation}
X^A = Y^A - D^A g,
\label{23}
\end{equation}

where $Y^A$ is divergence-free $D_A Y^A = 0$ (and thus a gauge redundancy) and $g$ is a scalar function on $S^2.$ A proof of \eqref{23} is given in the Appendix. The gauge aspects \eqref{21} are thus parametrized by two scalars on $S^2$ 

\begin{equation} \label{24}
\begin{split}
X &= f \\
X^A &= -D^A g
\end{split}
\end{equation}

and this parametrization is unique up to constant shifts in $g,$ which constitute gauge redundancies. As noted, $f$ describes $\mathfrak{bms}_4$-supertranslations. What is the meaning of $g$? Out of the excitations \eqref{24}, precisely the choice $\zeta=\zeta(X,X^A)$ with

\begin{equation} \label{25}
\begin{split}
X &= f \\
X^A &= -\frac{1}{r_S} D^A f
\end{split}
\end{equation}

keeps the induced metric on the event horizon invariant for an arbitrary scalar $f$ on $S^2.$ One has $\zeta|_{r=r_S}=f \partial_v.$ Due to these similarities with the ordinary $\mathfrak{bms}_4$-supertranslations at null infinity, the excitations \eqref{25} are identified as event horizon supertranslations. In the limit $r_S \to \infty$ the future event horizon tends to past null infinity and indeed the event horizon supertranslations \eqref{25} converge to the $\mathfrak{bms}_4$-supertranslations at past null-infinity. We arrive at the conclusion, that the degrees of freedom of a Schwarzschild black hole are given by $\mathfrak{bms}_4$-supertranslations and the event horizon supertranslations \eqref{25}. The latter contain a pure $\mathfrak{bms}_4$-supertranslation part. As these excitations reflect the degeneracy of the gravitational vacuum, we subtract them to obtain the candidates for the black hole microstates

\begin{equation} \label{26}
\begin{split}
X &= 0 \\
X^A &= -D^A g
\end{split}
\end{equation}

with a scalar function $g$ on $S^2.$ Thus, the asymptotic symmetries of the Schwarzschild solution $g_{ab}$ are enhanced by the $\mathcal{A}$-modes \eqref{26} with respect to the asymptotic symmetry algebra $\mathfrak{bms}_4$ present also in the case without event horizon. Already in \cite{Averin:2016ybl} the $\mathcal{A}$-modes were by this pure geometric reasoning (although in a different gauge) proposed as candidates for the microstates. It is nice to see, that a symplectic reasoning tends to the same answer. 

In addition, the $\mathcal{A}$-modes \eqref{26} violate the Bondi fall-off conditions as expected in chapter 2 for potential candidates for black hole microstates. That is, the set of data \eqref{6} is not enough to specify the excitations of $g_{ab}$ given by \eqref{7}. At the point $g_{ab}$ in phase space the gauge aspect $g$ provides additional Cauchy-data as it is a physical degree of freedom. 

To summarize, in this chapter we have analyzed the Hamiltonian phase space near the point $g_{ab}$ \eqref{7} \eqref{12} describing a Schwarzschild spacetime. Precisely, we analyze the tangent space $T_{g_{ab}} \Gamma$ of the phase space right at the point $g_{ab} \in \Gamma.$ Motivated by black hole uniqueness theorems/soft hair proposal, we further restricted to tangent vectors $h_{ab}$ that have the form of gauge transformations, i.e. that correspond to gauge excitations of $g_{ab}.$ Gauge-fixing to Bondi-gauge \eqref{13}, we constructed the Hamiltonian generators of these gauge excitations \eqref{19}. We inferred that all physical gauge excitations of $g_{ab}$ (i.e. those which are not gauge redundancies) are parametrized by the gauge aspects \eqref{24}. They consist of $\mathfrak{bms}_4$-supertranslations reflecting the degeneracy of gravitational vacua. In addition, there are $\mathcal{A}$-modes \eqref{26} violating Bondi fall-offs and thus giving rise to additional BCD in \eqref{6} as expected in chapter 2 for excitations describing microstates. Thus, we propose the $\mathcal{A}$-modes \eqref{26} to be responsible for black hole microstates of $g_{ab}.$ 

\section{Surface Charge Algebra}

In the last chapter, we figured out the surface degrees of freedom of a Schwarzschild black hole. They are elements of the tangent space at $g_{ab}$ describing gauge-fixed gauge excitations. In order to find their surface charge algebra, we need to make some technical considerations about how gauge-fixing takes place in the covariant phase space formalism. Let now $h_{ab} = \mathcal{L}_{\xi} g_{ab} \in T_{g_{ab}} \bar{\mathcal{F}}$ be a gauge excitation, which need not be gauge-fixed. That means, the vectorfield $\xi$ has not to be a residual gauge transformation with respect to Bondi-gauge.

The Hamiltonian generators \eqref{17} define linear forms on this tangent space

\begin{equation}
\delta H_{(X,X^A)}[h_{ab};g_{ab}] = -\frac{r_S^2}{16 \pi G} \oint_{\partial \Sigma} {d^2x \sqrt{\gamma} F_{rv}|_{r=r_S}},
\label{27}
\end{equation}

where

\begin{equation*}
\begin{split}
\left. F_{rv} \right|_{r=r_S} &= \zeta^A\left(\partial_r h_{Av} -\frac{2}{r_S}h_{Av}\right) +\zeta^v \left(-\frac{1}{r_S^2}D^A h_{Av} + \frac{1}{r_S^2}\partial_v {h^A}_A - \frac{2}{r_S}h_{vv} \right. \\
&- \left. \frac{1}{2r_S^3}{h^A}_A\right)
+ \partial_r \zeta^v h_{vv} +\frac{1}{r_S^2}D^A\zeta^v h_{vA} -\frac{1}{2r_S^2}\partial_v \zeta^v {h^A}_A + \zeta^r \left(\frac{2}{r_S}h_{vr} \right. \\
&+ \left. \frac{1}{r_S^3}{h^A}_A\right)
+ \frac{1}{2r_S^2}\partial_r \zeta^r {h^A}_A.
\end{split}
\end{equation*}

Here, $\zeta = \zeta(X,X^A)$ is of the form \eqref{13} with

\begin{equation} \label{28}
\begin{split}
X &= f \\
X^A &= -D^A g,
\end{split}
\end{equation}

with scalar functions $f,g$ on $S^2$ as found for the surface degrees of freedom in \eqref{24}. For an arbitrary vectorfield $\xi$ on the spacetime, the linear form \eqref{27} has for the gauge excitation $h_{ab} = \mathcal{L}_{\xi} g_{ab}$ the form

\begin{equation} \label{29}
\begin{split}
&\delta H_{(X,X^A)}[h_{ab},g_{ab}]
=-\frac{r_S^2}{16 \pi G} \oint_{\partial \Sigma} d^2x \sqrt{\gamma} \left(X^A r_S^2\gamma_{AB} \partial_r \partial_v \xi^B \right. \\
&+D_A X^A \left(\frac{1}{r_S}\xi^v -2\partial_r\xi^r-\frac{1}{r_S}\xi^r-\partial_v \xi^v -\frac{3}{2}D_B \xi^B\right) \\
&+X \left(-r_S\partial_r \partial_v D_B \xi^B +\frac{1}{r_S^2}D^2 \xi^v -\frac{2}{r_S}\partial_r D^2 \xi^r -\frac{2}{r_S^2}D^2 \xi^r \right. \\
&\left. \left.  -\frac{1}{r_S}D_B \xi^B-\frac{1}{r_S}\partial_v D^2 \xi^v -\frac{1}{r_S}D^2 D_B \xi^B \right) \right).
\end{split}
\end{equation}

Performing on the vectorfield $\xi^A$ on $S^2$ the decomposition \eqref{A1}

\begin{equation}
\xi^A = \tilde{\xi}^A + D^A h,
\label{30}
\end{equation}

where $\tilde{\xi}^A$ is divergence-free $D_A \tilde{\xi}^A = 0$ and $h$ is a scalar on $S^2,$ \eqref{29} is rewritten

\begin{equation} \label{31}
\begin{split}
&\delta H_{(X,X^A)}[h_{ab},g_{ab}]
=-\frac{r_S^2}{16 \pi G} \oint_{\partial \Sigma} d^2x \sqrt{\gamma} \left( \right. \\
&D_A X^A \left(-r_S^2\partial_r \partial_v h+\frac{1}{r_S}\xi^v -2\partial_r\xi^r-\frac{1}{r_S}\xi^r-\partial_v \xi^v -\frac{3}{2}D_B \xi^B\right) \\
&+X \left(-r_S\partial_r \partial_v D_B \xi^B +\frac{1}{r_S^2}D^2 \xi^v -\frac{2}{r_S}\partial_r D^2 \xi^r -\frac{2}{r_S^2}D^2 \xi^r \right. \\
&\left. \left.  -\frac{1}{r_S}D_B \xi^B-\frac{1}{r_S}\partial_v D^2 \xi^v -\frac{1}{r_S}D^2 D_B \xi^B \right) \right).
\end{split}
\end{equation}

If $\xi = \xi(Y,Y^A)$ is itself chosen to be an excitation of the surface degrees of freedom with gauge aspects $Y,Y^A$ (see \eqref{13} and \eqref{24}), we get as in \eqref{19}

\begin{equation} \label{32}
\begin{split}
&\delta H_{(X,X^A)}[h_{ab}=h_{ab}(Y,Y^A);g_{ab}] \\
&=-\frac{r_S^2}{16 \pi G} \oint_{\partial \Sigma} {d^2x \sqrt{\gamma} \left(D_A X^A \cdot \frac{1}{r_S}(1-D^2)Y
+X \cdot \frac{1}{r_S}(D^2-1)D_B Y^B\right).}
\end{split}
\end{equation}

That means, an arbitrary gauge-excitation $\xi$ (not satisfying Bondi-gauge) excites (up to zero-modes of the symplectic form) the gauge aspects $(Y,Y^A)$ determined by

\begin{equation} \label{33}
\begin{split}
&\frac{1}{r_S}(1-D^2)Y \\   
&= \left. -r_S^2\partial_r \partial_v h+\frac{1}{r_S}\xi^v -2\partial_r\xi^r-\frac{1}{r_S}\xi^r-\partial_v \xi^v -\frac{3}{2}D_B \xi^B \right|_{\partial \Sigma} \\
&\frac{1}{r_S}(D^2-1)D_B Y^B \\
&= -r_S\partial_r \partial_v D_B \xi^B +\frac{1}{r_S^2}D^2 \xi^v -\frac{2}{r_S}\partial_r D^2 \xi^r -\frac{2}{r_S^2}D^2 \xi^r \\
&\left. -\frac{1}{r_S}D_B \xi^B-\frac{1}{r_S}\partial_v D^2 \xi^v -\frac{1}{r_S}D^2 D_B \xi^B \right|_{\partial \Sigma}.
\end{split}
\end{equation}

The right hand side of \eqref{33} has to be evaluated at the coordinates $(v,r=r_S),$ where $\partial \Sigma$ is located. Since $D^2-1$ is an invertible operator on $S^2,$ \eqref{33} defines uniquely the gauge aspects $(Y,D_A Y^A)$ as functions on $S^2.$ The gauge excitation $\xi$ can excite additional degrees of freedom corresponding to shifts of other Cauchy-data in \eqref{6}. For example, $\xi$ can excite also radiative degrees of freedom describing radiation passing through the event horizon or null-infinity. To determine the correct shifts in phase space the symplectic form \eqref{29} has to be evaluated both also with respect to all others than the surface degrees of freedom $(X,X^A)$ and the location of $\partial \Sigma$ has to be varied across a whole Cauchy-surface. However, rather than doing a complete analysis of the phase space, we restrict ourselves to the surface degrees of freedom. Their excitations are are given (up to zero-modes of the linear forms \eqref{29}, i.e. up to gauge redundancies) by \eqref{33}.

In other words, \eqref{33} defines a projection operator, which maps the subspace of gauge-excitations $h_{ab} = \mathcal{L}_{\xi} g_{ab} \in T_{g_{ab}} \bar{\mathcal{F}}$ in the tangent space $T_{g_{ab}} \bar{\mathcal{F}}$ to an excitation in $T_{g_{ab}} \Gamma$ of the surface degrees of freedom with gauge aspects $(Y,D_A Y^A).$ 

Let now

\begin{equation} \label{34}
\begin{split}
\xi_1(X_1,X_1^A) &= X_1\partial_v -\frac{1}{2}\left(rD_A X_1^A+D^2 X_1\right)\partial_r +\left(X_1^A+\frac{1}{r}D^AX_1\right) \partial_A \\
\xi_2(X_2,X_2^A) &= X_2\partial_v -\frac{1}{2}\left(rD_A X_2^A+D^2 X_2\right)\partial_r +\left(X_2^A+\frac{1}{r}D^AX_2\right) \partial_A
\end{split}
\end{equation}

be two gauge excitations with gauge aspects

\begin{equation} \label{35}
\begin{split}
X_1 &= f_1 \\
X_1^A &= -D^A g_1 \\
X_2 &= f_2 \\
X_2^A &= -D^A g_2.
\end{split}
\end{equation}

What are the gauge aspects (according to the projector \eqref{33}) of the Lie-bracket $\left[ \xi_1,\xi_2\right]$?

We have

\begin{equation}
\left[ \xi_1,\xi_2\right]^v = X_1^A D_A X_2 - X_2^A D_A X_1
\label{36}
\end{equation}

as well as

\begin{equation} \label{37}
\begin{split}
\left[ \xi_1,\xi_2\right]^r &= r\left(-\frac{1}{2}X_1^AD_AD_BX_2^B +\frac{1}{2}X_2^AD_AD_BX_1^B\right) \\
&+ \left( \frac{1}{4}D^2X_1D_BX_2^B -\frac{1}{2}X_1^AD_AD^2X_2 -\frac{1}{2}D^AX_1D_AD_BX_2^B \right. \\
&- \left. \frac{1}{4}D^2X_2 D_BX_1^B +\frac{1}{2}X_2^AD_AD^2X_1 +\frac{1}{2}D^AX_2D_AD_BX_1^B \right) \\
&+ \frac{1}{r}\left( -\frac{1}{2}D^A X_1D_AD^2X_2 +\frac{1}{2}D^AX_2D_AD^2X_1 \right)
\end{split}
\end{equation}

and

\begin{equation} \label{38}
\begin{split}
\left[ \xi_1,\xi_2\right]^A &= \left(X_1^BD_BX_2^A - X_2^BD_BX_1^A\right) \\
&+\frac{1}{r}\left(\frac{1}{2}D_BX_1^BD^AX_2 - \frac{1}{2}D_BX_2^BD^AX_1 \right. \\
&+X_1^BD_BD^AX_2 - X_2^BD_BD^AX_1 \\
&+ \left. D^BX_1D_BX_2^A - D^BX_2D_BX_1^A \right) \\
&+\frac{1}{r^2}\left(\frac{1}{2}D^2X_1D^AX_2 - \frac{1}{2}D^2X_2D^AX_1 \right. \\
&+ \left. D^BX_1D_BD^AX_2 - D^BX_2D_BD^AX_1\right).
\end{split}
\end{equation}

From this, we infer for the gauge aspects

\begin{equation}
(Y,D_A Y^A) = (\hat{f}, -D^2 \hat{g})
\label{39}
\end{equation}

of the Lie-bracket $\left[ \xi_1,\xi_2\right]$ from \eqref{33}

\begin{equation} \label{40}
\begin{split}
&\frac{1}{r_S}(1-D^2)Y \\
&= \frac{1}{r_S^2} \left( -\frac{5}{4}D^AX_1D_AD^2X_2 + \frac{5}{4} D^AX_2D_AD^2X_1 \right) \\
&+\frac{1}{r_S}\left(X_1^A D_AX_2 - X_2^AD_AX_1 \right. \\
&-X_1^AD_AD^2X_2 + X_2^AD_AD^2X_1 \\
&-\frac{1}{2}D_AX_1^AD^2X_2 + \frac{1}{2}D_AX_2^AD^2X_1 \\
&+ \left. \frac{1}{4}D_AD_BX_1^BD^AX_2 - \frac{1}{4}D_AD_BX_2^BD^AX_1 \right)
\end{split}
\end{equation}

and

\begin{equation} \label{41}
\begin{split}
&\frac{1}{r_S}(D^2-1)D_B Y^B \\
&= \frac{1}{r_S}(D^2-1)D_A\left(X_1^AD_BX_2^B - X_2^AD_BX_1^B\right) \\
&+\frac{1}{r_S^2}\left(-X_1^BD_BD^2X_2 - \frac{1}{2}D_BX_1^BD^2X_2 + \frac{1}{2}D_AD_BX_1^BD^AX_2 \right. \\
&+X_2^BD_BD^2X_1 + \frac{1}{2}D_BX_2^BD^2X_1 - \frac{1}{2}D_AD_BX_2^BD^AX_1 \\
&+D^2 \left(-\frac{1}{2}D^AX_2D_AD_BX_1^B + X_1^AD_AX_2 \right. \\
&+\left. \left. \frac{1}{2}D^AX_1D_AD_BX_2^B - X_2^AD_AX_1\right) \right) \\
&+\frac{1}{r_S^3}\left( \frac{1}{2}D_AD^2X_1D^AX_2 - \frac{1}{2}D_AD^2X_2D^AX_1 \right. \\
&+\left.D^2\left( \frac{1}{2}D_AD^2X_1D^AX_2 - \frac{1}{2}D_AD^2X_2D^AX_1 \right) \right).
\end{split}
\end{equation}

On the surface degrees of freedom \eqref{35}, the conventional spacetime Lie-bracket is realized through the algebra 

\begin{equation} \label{42}
\begin{split}
&(1-D^2)\hat{f} \\
&= \frac{1}{r_S} \left( -\frac{5}{4}D^Af_1D_AD^2f_2 + \frac{5}{4} D^Af_2D_AD^2f_1 \right) \\
&+\left(-D^Ag_1 D_Af_2 + D^Ag_2D_Af_1 \right. \\
&+D^Ag_1D_AD^2f_2 - D^Ag_2D_AD^2f_1 \\
&+\frac{1}{2}D^2g_1D^2f_2 - \frac{1}{2}D^2g_2D^2f_1 \\
&- \left. \frac{1}{4}D_AD^2g_1D^Af_2 + \frac{1}{4}D_AD^2g_2D^Af_1 \right)
\end{split}
\end{equation}

and

\begin{equation} \label{43}
\begin{split}
&(1-D^2)D^2 \hat{g} \\
&= (D^2-1)D_A\left(D^Ag_1D^2g_2 - D^Ag_2D^2g_1\right) \\
&+\frac{1}{r_S}\left(D^Bg_1D_BD^2f_2 + \frac{1}{2}D^2g_1D^2f_2 - \frac{1}{2}D_AD^2g_1D^Af_2 \right. \\
&-D^Bg_2D_BD^2f_1 - \frac{1}{2}D^2g_2D^2f_1 + \frac{1}{2}D_AD^2g_2D^Af_1 \\
&+D^2 \left(\frac{1}{2}D^Af_2D_AD^2g_1 - D^Ag_1D_Af_2 \right. \\
&-\left. \left. \frac{1}{2}D^Af_1D_AD^2g_2 + D^Ag_2D_Af_1\right) \right) \\
&+\frac{1}{r_S^2}\left( \frac{1}{2}D_AD^2f_1D^Af_2 - \frac{1}{2}D_AD^2f_2D^Af_1 \right. \\
&+\left.D^2\left( \frac{1}{2}D_AD^2f_1D^Af_2 - \frac{1}{2}D_AD^2f_2D^Af_1 \right) \right).
\end{split}
\end{equation}

It is known, that the Hamiltonian generators form a representation (with respect to the Poisson-bracket) of the Lie-algebra of symplectic symmetries up to central extensions. That is,

\begin{equation}
\left\{ H_X,H_Y \right\} = H_{[X,Y]} + K_{X,Y}
\label{44}
\end{equation}

for symplectic symmetries $X,Y$ and their generators $H_X, H_Y.$ The central extension $K_{X,Y}$ is a c-number and $[X,Y]$ is the Lie-bracket of $X$ and $Y$ as vectorfields on the phase space. If $X=\delta_{\xi_1}$ and $Y=\delta_{\xi_2}$ are gauge transformations, we assume that \eqref{44} takes on-shell the form

\begin{equation}
\left\{ H_{\xi_1}, H_{\xi_2}\right\} = H_{[\xi_1,\xi_2]} + K_{\xi_1,\xi_2}
\label{45}
\end{equation}

with $[\xi_1,\xi_2]$ being the Lie-bracket of vectorfields on the spacetime manifold. That means, on shell $[X,Y]=\delta_{[\xi_1,\xi_2]}$ up to gauge redundancies.\footnote{Although \eqref{44} is often used in the form \eqref{45} \cite{Carlip:1999cy} \cite{PerryTalk} \cite{Koga:2001vq}, we do not know a proof of that. We further comment on this assumption in the next chapter. For now, in this chapter we justify the use of \eqref{45} by being able to reproduce known results.}

Choosing in \eqref{45} for the gauge transformations the surface degrees of freedom \eqref{34}, we get

\begin{equation}
\left\{ H_{(X_1,X_1^A)}, H_{(X_2,X_2^A)}\right\} = H_{(Y,Y^A)} + K_{(X_1,X_1^A),(X_2,X_2^A)}.
\label{46}
\end{equation}

Remembering $\left\{ H_{(X_1,X_1^A)}, H_{(X_2,X_2^A)}\right\} = \delta_{(X_2,X_2^A)} H_{(X_1,X_1^A)}$ we get the central term from \eqref{19}

\begin{equation} \label{47}
\begin{split}
K_{(X_1,X_1^A),(X_2,X_2^A)} &= \frac{r_S}{16 \pi G} \oint_{\partial \Sigma} {d^2x \sqrt{\gamma} X_1(-D^2+1)D_AX_2^A} \\
&-\frac{r_S}{16 \pi G} \oint_{\partial \Sigma} {d^2x \sqrt{\gamma} X_2(-D^2+1)D_AX_1^A} \\
&-H_{(Y,Y^A)}[g_{ab}].
\end{split}
\end{equation}

Hamiltonian generators are determined only up to a constant. We use this freedom to set all surface charges to $0$ at the reference solution $g_{ab}$

\begin{equation}
H_{(X,X^A)}[g_{ab}] = 0.
\label{48}
\end{equation}

This choice fixes uniquely all generators and the central terms  \eqref{47}.

To summarize, for the surface degrees of freedom \eqref{24} of a Schwarzschild black hole, the surface charge algebra is given by 

\begin{equation}
\left\{H_{f_1,g_1},H_{f_2,g_2}\right\} = H_{\hat{f},\hat{g}} + K(f_1,g_1;f_2,g_2).
\label{49}
\end{equation}

Here, the gauge aspects $\hat{f}$ and $\hat{g}$ are given by the algebra \eqref{42} \eqref{43} and the central term follows from \eqref{47} (with the choice \eqref{48})

\begin{equation} \label{50}
\begin{split}
K(f_1,g_1;f_2,g_2) &= \frac{r_S}{16 \pi G} \oint_{\partial \Sigma} {d^2x \sqrt{\gamma} f_1(D^2-1)D^2g_2} \\
&-\frac{r_S}{16\pi{G}} \oint_{\partial \Sigma} {d^2x \sqrt{\gamma} f_2(D^2-1)D^2g_1}.
\end{split}
\end{equation}

We comment on some implications of this algebra. First, we have for the choice $f_1=r_S, g_1=0$ and $f_2=f, g_2=g$ the bracket 

\begin{equation}
\left\{H_{r_S,0},H_{f,g}\right\} = 0.
\label{51}
\end{equation}

The charge $H_{r_S,0}$ is (up to constant shift set by \eqref{48} and normalization) equal to the ADM-energy subtracted of by the energy passing through future null infinity and the portion of the event horizon between the location of $\partial \Sigma$ and the horizon's future end point. Thus, if there is no radiation passing through these regions, $H_{r_S,0}$ coincides with the ADM-energy. $\eqref{51}$ then states that the surface degrees of freedom are gapless excitations, i.e. they keep the ADM-energy invariant. They provide soft black hole hair. As mentioned, the $\mathfrak{bms}_4$-supertranslations $f$ reflect degeneracy of the gravitational vacuum. The $\mathcal{A}$-modes $g$ are the gapless Bogoliubov-modes associated with the criticality of the Schwarzschild black hole. 

Furthermore, as a consistency check, we find that the Poisson-bracket between event horizon supertranslations \eqref{25} (i.e. choosing $g_i=\frac{1}{r_S}f_i$ for $i=1,2$ and arbitrary $f_i$ in \eqref{49}) vanishes. This is in agreement with \cite{Donnay:2016ejv} \cite{Donnay:2015abr}.

We have identified the surface degrees of freedom of a Schwarzschild black hole as the gauge aspects, which are functions on $S^2.$ The algebra with respect to the Poisson-bracket of the gauge aspects is given by \eqref{49}. We thus arrived at a lower dimensional theory describing part of the phase space near the Schwarzschild solution $g_{ab}.$ Thus, we have found a new and concrete realization of the holographic principle \cite{tHooft:1993dmi} \cite{Susskind:1994vu} for the case of a Schwarzschild black hole. 

\section{Assumptions, Limitations and Outlook}

After having found a dual theory for the Schwarzschild black hole, it is interesting to analyze its consequences. However, we want to warn that in our path, we made several assumptions. These assumptions may cause corrections to our results. In this chapter, we want to list these assumptions and give an outlook. Further investigation of these issues will be left for future research. 

\subsection{Choice of symplectic form, integrability vs. Gibbons-Hawking-York term}

Given the Lagrangian of a theory, the covariant phase space formalism starts by prescribing a presymplectic potential. Unfortunately, this prescription is affected by adding a boundary term to the action and has a further ambiguity on its own (see \cite{Seraj:2016cym}). These ambiguities affect the definition of the presymplectic form and therefore also the Hamiltonian generators. As commonly done in the literature, we used in our derivations of formulas like \eqref{19} the canonical presymplectic potential as derived from the Einstein-Hilbert action.

On the other hand, in the Hamiltonian approach (see \cite{Troessaert:2013fba}) any ambiguity in the definition of the Hamiltonian generators is fixed (of course up to a constant) by the requirement of differentiability in the sense of Regge-Teitelboim \cite{Regge:1974zd}. Having found a candidate for a Hamiltonian generator of a symplectic symmetry, a suited boundary term has to be added to make the generator a differentiable functional over phase space. This fixes any ambiguity. 

Having a theory with a well-defined action, that means, an action that is added a suited boundary term to ensure Regge-Teitelboim differentiability in the variational principle, there is the following version of Noether's theorem incoorporating boundary effects:

For a symmetry of a well-defined action, the canonical Noether-procedure assigns a charge which is a differentiable Hamiltonian generator of that symmetry (see \cite{Troessaert:2013fba} for the details). 

The derivation of black hole entropy in \cite{Gibbons:1976ue} using Euclidean methods suggests that variation of the Gibbons-Hawking-York boundary term $S_{GHY}$ vanishes

\begin{equation}
\left. \delta_\xi S_{GHY} \right|_{g_{ab}} = 0
\label{52}
\end{equation}

for the physical gauge excitations $\xi$ of the black hole geometry $g_{ab}$ that are responsible for the microstates. That means first, that for the construction of the Hamiltonian generators of the $\xi$s, the boundary term in the action does not affect the presymplectic potential. Second, the above Noether-theorem guarantees the existence of differentiable Hamiltonian generators constructed by the canonical Noether-procedure. 

In summary, the canonical choice of the presymplectic potential (that we used throughout) is justified for the problem. However, it has to be checked that for our surface degrees of freedom \eqref{52} is indeed satisfied

\begin{equation}
\left. \delta_{f,g} S_{GHY} \right|_{g_{ab}} = 0
\label{53}
\end{equation}

for all gauge aspects $f,g$ and the reference metric $g_{ab}.$

Note that the above Noether-theorem also guarantees integrability of the Hamiltonian generators \eqref{19} over a suited region in phase space near $g_{ab}.$ Note also that over the last chapter, we assumed integrability, which is in general not guaranteed. 

Our physical interpretation of \eqref{53} is that gauge excitations $f,g$ do not excite gravitational radiation passing through boundaries of spacetime. It was already noted in \cite{Wald:1999wa} that integrability of Hamiltonian generators is spoiled by flux terms. 

\subsection{Lie-bracket vs. surface deformation bracket}

As noted in the last chapter, the algebra $\eqref{44}$ was assumed to take the form \eqref{45} on-shell. Although \eqref{45} is often used \cite{Carlip:1999cy} \cite{Koga:2001vq} \cite{PerryTalk}, we are not aware of a general proof. In the Hamiltonian approach \cite{Brown:1988am} a known result states that for spacetime vectorfield $\xi_1, \xi_2$ one has the relation

\begin{equation}
\left\{ \Gamma_{\xi_1}, \Gamma_{\xi_2} \right\} = \Gamma_{\{\xi_1,\xi_2\}_{\text{SD}}} + K_{\xi_1,\xi_2}
\label{54}
\end{equation}

if differentiable Hamiltonian generators $\Gamma_{\xi_1}, \Gamma_{\xi_2}$ are existent. Here, $\{\xi_1,\xi_2\}_{\text{SD}}$ is the surface deformation bracket which is in general different from the Lie-bracket $[\xi_1,\xi_2].$ The difference is calculated in \cite{Brown:1988am} and it is argued why it often happens (but not has to happen) that on-shell

\begin{equation}
\Gamma_{\{\xi_1,\xi_2\}_{\text{SD}}} =  \Gamma_{[\xi_1,\xi_2]}.
\label{55}
\end{equation}

\eqref{55} has to be checked and this was the assumption made in the derivation of the surface charge algebra in the last chapter.

\subsection{Sugawara-construction of 2D stress-tensor and entropy counting}

In the last chapter, we found a lower-dimensional theory on $S^2$ with the gauge aspects as degrees of freedom and their Poisson-brackets given by \eqref{49}. This theory describes part of the phase space near the Schwarzschild solution $g_{ab}.$ Note that so far, we did not specify how the word ``near'' has to be understood.

Strictly speaking, we performed our calculations right at the reference point $g_{ab}$ in phase space and in the tangent space thereof (see formulas like \eqref{19}). As explained in chapter 5.1 the algebra \eqref{49} is derived under the assumption of integrability. That is, for the generators of gauge aspects, \eqref{15} defines a $1$-form $\delta H_{f,g}$ over phase space $\Gamma$ which can over a suited region $S \subseteq \Gamma$ be integrated to obtain generators $H_{f,g}$ satisfying the algebra \eqref{49} over this region $S \subseteq \Gamma.$ Our analysis in $T_{g_{ab}} \bar{\mathcal{F}}$ was powerful enough to obtain the algebra \eqref{49}. However, only at the point $g_{ab},$ we know how the excitation of the gauge aspects generated by $H_{f,g}$ looks like (see \eqref{14} with \eqref{24}). The action of $H_{f,g}$ at other points in $S,$ we do not know in general. Of course, the residual gauge transformations at other points in $S$ look different than in \eqref{13}. Neither, we know how large the region $S \subseteq \Gamma$ is. We want to argue for a reasonable $S$ by asking what the theory obtained actually describes?

Since we showed, that the gauge aspects are gapless excitations of a Schwarzschild black hole, $S$ should contain these points. As already explained in chapter 1, this scale invariance suggests that our two-dimensional theory is a conformal field theory. This Schwarzschild/CFT-correspondence would then deliver a two-dimensional CFT which describes the part of the phase space $S$ of the full four-dimensional Einstein-gravity. $S$ at least contains the gapless excitations of the Schwarzschild black hole. 

A conformal anomaly (as suggested by the appearance of central terms in \eqref{49}) would then reflect the quantum mechanical lifting of gapless modes by $\frac{1}{S}$-corrections as explained in the introduction.

If the dual theory of the last chapter is indeed conformally invariant, it has to posess a 2D stress-tensor with the Virasoro-algebra being compatible with \eqref{49}. Since we know the algebra \eqref{49}, it is natural to search for the stress-tensor via a Sugawara-construction. That is, we construct the Virasoro-generators out of the surface degrees of freedom under the requirement of validity of the Virasoro-algebra. As an ansatz for the Virasoro-generators, we motivate ourselves with the cases of the Brown-Henneaux analysis \cite{Brown:1986nw} or the case of extremal Kerr/CFT \cite{Guica:2008mu} \cite{Bredberg:2011hp}. There, the Virasoro-generators themselves are the generators of suited gauge transformations. Following 5.2, we search for spacetime vectorfields satisfying a Witt-algebra with respect to the Lie-bracket. The associated generators from the gauge aspects (obtained with the projection operator \eqref{33}) then satisfy via \eqref{49} a Virasoro-algebra and thus are candidates for the Virasoro-generators building the stress-tensor. 

To this end, we define the spacetime vectorfields

\begin{align}
\xi_n^a = 
\begin{pmatrix}
\xi_n^v \\
\xi_n^r \\
\xi_n^\theta \\
\xi_n^\phi
\end{pmatrix}
=
\begin{pmatrix}
2r_SA\left(1-\frac{A}{A+B}\left(1+\frac{in}{A}\right) \right) \\
r_S\left(\frac{r_S}{r}-1\right)in \\
0 \\
\frac{A}{A+B}\left(1+\frac{in}{A}\right)
\end{pmatrix}
e^{\frac{in}{2r_SA}v}e^{in\phi}
\label{56}
\end{align}

and

\begin{align}
\bar\xi_n^a = 
\begin{pmatrix}
\bar\xi_n^v \\
\bar\xi_n^r \\
\bar\xi_n^\theta \\
\bar\xi_n^\phi
\end{pmatrix}
=
\begin{pmatrix}
-\frac{2r_SAB}{A+B}\left(1+\frac{in}{B}\right) \\
-inr_S \left(1-\frac{r_S}{r}\right) \\
0 \\
\frac{B}{A+B}\left(1+\frac{in}{B}\right)
\end{pmatrix}
(-1)^{\frac{in}{B}} e^{\frac{inr^*}{r_SB}} e^{-\frac{inv}{2r_SB}} e^{in\phi}
\label{57}
\end{align}

for $n \in \mathbb{Z}.$ The vectorfields are given in infalling Eddington-Finkelstein coordinates used in chapter 2. The constants $A, B \in \mathbb{R}$ are arbitrary. We then have $(\xi_n^a)^* = \xi_{-n}^a$ and $(\bar\xi_n^a)^* = \bar\xi_{-n}^a.$ They fulfill two copies of the Witt-algebra

\begin{equation} \label{58}
\begin{split}
\left[\xi_m, \xi_n\right] &= -i(m-n) \xi_{m+n} \\
\left[\bar\xi_m, \bar\xi_n\right] &= -i(m-n) \bar\xi_{m+n}.
\end{split}
\end{equation}

The choice is motivated by similar vectorfields appearing in Carlip's approach to entropy counting in \cite{Carlip:1998wz} \cite{Park:2001zn} but changed in such a way as to satisfy Witt-algebra \eqref{58} and treat future and past horizon equally. Similar vectorfields appear in \cite{PerryTalk}. Let $(f_n,g_n)$ be the associated gauge aspects to \eqref{56}. Furthermore, let

\begin{equation}
H_n := H_{(f_n,g_n)}
\label{59}
\end{equation}

be the associated Hamiltonian generators under the choice \eqref{48} $H_n[g_{ab}]=0$ for the fixed reference solution $g_{ab}.$ Since $H_n$ has dimension of an action, we can define dimensionless generators

\begin{equation}
\hbar L_n := H_n + \frac{r_S^2}{4G} \frac{2A^2B+B-A}{(A+B)^2} \delta_n
\label{60}
\end{equation}

for $n \in \mathbb{Z}$ and with $\delta_n = \delta_{n,0}$ being the Kronecker delta.

Computing the central terms from the algebra \eqref{49} and under the assumptions of this chapter, we get the classical Virasoro-algebra

\begin{equation}
\left\{L_m,L_n\right\} = -\frac{i}{\hbar} (m-n) L_{m+n} -\frac{i}{\hbar^2}\frac{r_S^2}{2G}\frac{B-A}{(A+B)^2}m(m^2-1)\delta_{m+n}.
\label{61}
\end{equation}

Canonical quantization yields a Virasoro-algebra with (using standard conventions)

\begin{equation} \label{62}
\begin{split}
c &= \frac{6r_S^2}{\hbar G} \frac{B-A}{(A+B)^2} \\
L_0[g_{ab}] &= \frac{r_S^2}{4 \hbar G} \frac{2A^2B+B-A}{(A+B)^2}.
\end{split}
\end{equation}

We note that our computation of surface charges in \eqref{27} and thus of gauge aspects use $\partial \Sigma$ to be located on the future event horizon at a particular time $v.$ Whereas the gauge aspects of \eqref{56} $(f_n,g_n)$ depend on the choice of $v,$ the result \eqref{62} does not. Unfortunately, the computation of gauge aspects of \eqref{57} contains divergences. This is due to the fact, that whereas \eqref{56} is regular at the future event horizon, \eqref{57} is at the past event horizon but are singular vice versa. Performing the computation of the gauge aspects $(\bar{f}_n,\bar{g}_n)$ of \eqref{57} at the past event horizon, the anti-chiral analog of \eqref{62} $\bar{c}, \bar{L}_0$ does not depend on the location of $\partial \Sigma$ and thus the limit of taking $\partial \Sigma$ to the bifurcation of the horizons is for the evaluation of the Virasoro-algebras well-defined. Unfortunately, the projection formulas \eqref{33} are not suited to determine the anti-chiral gauge aspects $(\bar{f}_n,\bar{g}_n).$ This is due to the fact, that their derivation has to be refined in that (working in the advanced Bondi-gauge) the limit where $\partial \Sigma$ goes to the past horizon has to be taken carefully. We note that these issues are under current investigation. The hope then is, that counting the degeneracy with the Cardy-formula matches Bekenstein-Hawking entropy. However, there must be a finite result for the anti-chiral analog of \eqref{62} as we could have also performed the calculation in retarded Bondi-gauge. The gauge aspects would then have to be matched by a similar matching condition as the one in \cite{Strominger:2013jfa}.

Wheras there are still issues under current investigation, our approach sheds new light on Carlip's approach to a microcanonical counting of entropy \cite{Carlip:1998wz} \cite{Carlip:1999cy}. In Carlip's approach, the choice of vectorfields giving rise to Virasoro-algebra seems ad-hoc. The near-horizon asymptotic symmetry algebra has to be unnaturally reduced to yield a Virasoro-algebra with central terms for the generators \cite{Koga:2001vq}. In our approach, such a reduction is first due to dividing out zero-modes by projecting arbitrary gauge excitations onto the surface degrees of freedom via \eqref{33}. That is, different gauge excitations can correspond to the same excitations of the gauge aspects. Second, only the very special generators \eqref{60} correspond to Virasoro-generators out of the full set of generators of surface degrees of freedom. 

On the other hand, note that in \eqref{62} $r_S$ is the Schwarzschild-radius of the reference solution $g_{ab}.$ It is a fixed parameter for our dual theory. Also note the appearance of the two arbitrary parameters $A,B$ in \eqref{62}. Such an ambiguity was already present in Carlip's approach, although it was canceled in the entropy counting giving consistent result. This ambiguity reflects the fact that Hamiltonian generators are only defined up to constant. Had we chosen in \eqref{48} a different reference solution, we would have obtained a different theory \eqref{49} with other central terms and this would affect the associated Virasoro-algebra. This ambiguity is reflected in the presence of the parameters $A,B$ in \eqref{62}.

After all, it is tempting to expect that $S \subseteq \Gamma$ covers the whole Kerr-family. That is, we conjecture our dual theory describes part of the phase space containing the Kerr-family and its gapless excitations. Such a Kerr/CFT-correspondence was already conjectured in \cite{Castro:2010fd} from the study of scattering off a non-extremal Kerr black hole. Comparing with \eqref{56} \eqref{57} the Virasoro-modes $L_0, \bar{L}_0$ ``measure'' the mass and angular momentum parameter of a particular Kerr black hole.

Whether this is the right way to think about the problem has still to be understood. We have given an outlook of what is currently at our investigation.

\section*{Appendix: Proof of \eqref{23}}

In this chapter, we want to prove that a vectorfield $X^A$ on $S^2$ has a Helmholtz-Hodge decomposition

\begin{equation}
X^A = Y^A + D^A f,
\label{A1}
\end{equation}

where $Y^A$ is a divergence-free vectorfield on $S^2$ $D_A Y^A = 0$ and $f$ is a scalar function on $S^2.$ Proving \eqref{A1}, we have proven \eqref{23}.

Let $X^A$ be a vectorfield on $S^2.$ According to the Hodge-decomposition, we can write the $1$-form $X_A$ as

\begin{equation}
X = df + \delta \beta + \gamma,
\label{A2}
\end{equation}

where $f$ is a scalar on $S^2,$ $\beta$ is a $2$-form on $S^2$ and $\gamma$ is a harmonic $1$-form. $d$ denotes the exterior derivative and $\delta$ the codifferential. On $S^2,$ there are no harmonic $1$-forms, since the first de Rham cohomology-group vanishes. Thus, $\gamma = 0.$ Defining the vectorfield

\begin{equation}
Y^A = (\delta \beta)^A
\label{A3}
\end{equation}

\eqref{A1} follows immediately from \eqref{A2}

\begin{equation*}
X^A = D^A f + Y^A,
\end{equation*}

if we can show $D_A Y^A = 0.$ For a generic vectorfield $V^A,$ we have for the associated $1$-form $V_A$

\begin{equation}
\delta V = -*d(*V) = -**(D_A V^A) = -D_A V^A.
\label{A4}
\end{equation}

Using this identity, we conclude

\begin{equation*}
-D_A Y^A = \delta Y = \delta^2 \beta = 0.
\end{equation*}

$Y^A$ is indeed divergence-free and this shows \eqref{A1}.

\section*{Acknowledgements}
We thank Alexander Gußmann for many discussion on this and other topics in physics and proofreading the work. We thank Gia Dvali, Cesar Gomez, Dieter Lüst for discussions. We also thank Matthias Wink for discussions on Hodge theory.

\end{document}